# A Manga-Driven System Requirements Development PBL Exercise

Yasushi Tanaka
Nara Institute of Science and Technology,
Osaka University of Arts,
K-plus Solutions Co. Ltd.
Kobe, Hyogo, Japan
yasushi-tanaka@is.naist.jp

Hajimu Iida
Nara Institute of Science and Technology
Ikoma, Nara, Japan
iida@itc.naist.jp

Yasuhiro Takemura
Osaka University of Arts
Osaka, Japan
yasuhi-t@osaka-geidai.ac.jp

## ABSTRACT

We conducted a Project-Based Learning (PBL)-type exercise incorporating Japanese cartoon ("manga") techniques into Requirements Development (RD) processes. Manga has established techniques, such as those for character setting and story development, that we thought are also valid for RD processes. Using this manga-driven method, students were able to clarify high-level project goals early in the development life-cycle, and succeeded in defining high quality and unique system ideas.

## 1 INTRODUCTION

In planning this exercise to prepare students for upstream software development, we had three objectives. First was to introduce students to the Requirements Development (RD) process. Added to Capability Maturity Model Integration (CMMI) in 2000 [1]. 1, the RD process is now considered an integral part of software engineering. RD is a technique for eliciting customer needs and translating those needs into product requirements. Software engineers are required to know not only "how" but also "what" to develop in software. However, the "what" part of development is still little taught in most software engineering courses.

The second objective was to bring interdisciplinary work into the classroom. When an enterprise develops a new product, it is usually with a team of specialists from different fields, such as marketing, design, engineering, and manufacturing. We envisioned a collaborative exercise with engineering students and students in design or arts fields working together to reflect a real-world setting.

The third objective was to explore a novel technique that would improve the RD process when it comes to building the right product. This technique would complement other approaches, such as Design Thinking [2], UX Design [3], and Business Analysis [4].

To meet these objectives, we decided to create a PBL-type exercise focusing on upstream development. Cooperation between Nara Institute of Science and Technology (NAIST), a specialized science and technology graduate school, and the Character Creative Arts Department at Osaka University of Arts (OUA) enabled us to create a mixed class profile of both engineering and arts students as shown in Fig. 1. Furthermore, participation of manga majors from OUA gave us the idea of using manga in software development. In this paper, we explain the manga technique we experimented with in comparison to conventional methods used in RD processes, and describe the contents of this exercise, which started in 2014.

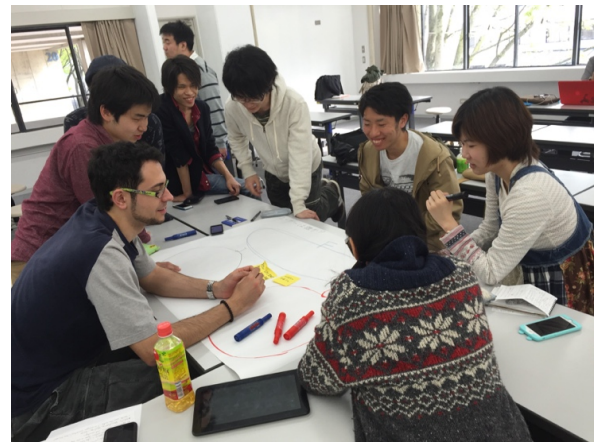

**Fig. 1: A mixed course with students of information science and students of manga majors.**

## 2 MANGA AND RD PROCESS METHODS

Different approaches have originated various methods for RD processes, like Use Case and User Story [5] for instance. Also, some have used Storyboards to guide the definition and validation of new design concepts [6]. Furthermore, the use of Iterative and Incremental Development is also an important aspect in applying the RD process to upstream software development.

Similar to Storyboard, the Manga technique provides a useful and simple communication tool between engineers and users to ensure building "the right thing". However, sometimes storyboards tend to focus too much on interfaces and not enough on the story's context. The Manga technique, with its story composition requirements, should avoid that pitfall.

### 2.1 Use Case, Storyboard, and Iterative and Incremental Development

*2.1.1 Use Case.* Use Case is a scenario driven approach. With Use Case, engineers can focus on the important parts of requirements. Furthermore, the quality assurance team can test the final product using the same scenario as the implemented concept. By investigating a Use Case, engineers can understand how the product achieves a business goal and a user's purpose. In addition, an appropriately defined Use Case can gather requirements for the middle layer such as the class and object model. Another advantage of Use Case is that the scenario is written in a natural language. Therefore, a user can understand and evaluate the meaning and discuss it.

*2.1.2 Storyboard.* If Use Case can be described as a synopsis of a movie, Storyboard is a script. While Use Case was born in the background of Software Engineering, Storyboard comes from Design Thinking and UX Design. There are no established standards in Storyboard, and various applications of the method have been tried [6]. In many cases, a specific user interface and hardware are drawn on the Storyboard, which effectively and clearly informs the design concept.

*2.1.3 Iterative and Incremental Development.* Iterative and Incremental Development is a method where users can instruct the implementation work at short intervals, or provide feedback on functionality. By combining Iterative and Incremental Development with Use Case and Storyboard, an engineer and a user can review a project and elucidate shortages in the requirements at an early stage in the development lifecycle.

### 2.2 Manga Composition Techniques

Below are the two important manga techniques we gathered from a professional cartoonist who teaches at OUA.

*2.2.1 Character Setting.* The most important technique in manga creation is the character setting. The professional cartoonist insisted that the quality of manga is determined not by storylines, but by the characters. In a novel, many people forget the name of the protagonist, but remember the plot. On the other hand, in a manga, many forget the story, but the image and the name of the hero remains vividly in a reader's memory. If a character is drawn concretely in manga, then the reader can identify with the character and participate in that story.

There are similar methods that give insights to users, such as the Empathy Map [7] and Persona [8-9]. The Empathy Map gives an understanding of users by mapping the user experience, such as what users are watching, saying, listening, and thinking. On the other hand, the Persona method describes users with a few personal characteristics, such as age, gender, educational level, and occupation. Based on research, it is used in UX Design and describes the user in a narrative form.

A major difference between these techniques and the manga technique is whether or not a picture can be drawn. To draw a manga character, the author must determine the "external features" of the character in detail, such as age, height, weight, appearance, way of speaking, laughter, and even how they cry. In addition, for manga, the character's relationship to his family and friends must also be considered. In this exercise, as shown in Fig. 2, student in information technology draw classmates' portraits to learn about character setting.

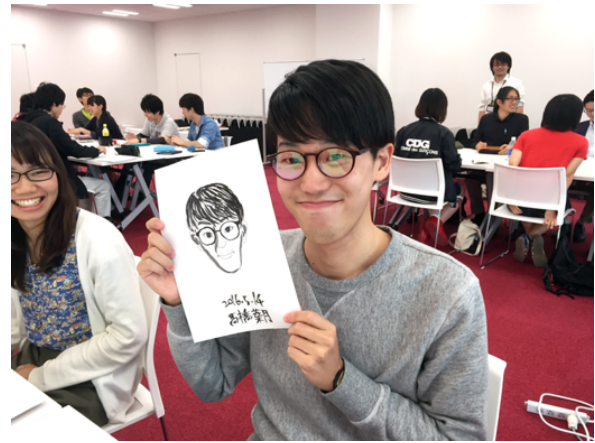

**Fig. 2: Student in Information Technology draw classmates' portraits to learn about Character Setting.**

*2.2.2 Story Composition.* A manga story uses the common East Asian narrative structure of setup, complications, twist, and climax, or "Ki-Sho-Ten-Ketsu".

More specifically, techniques of story composition in manga are as follows:

- Setup (Ki): Define a character
- Complications (Sho): Draw the premise of the story
- Twist (Ten): Draw the main character's dilemma
- Climax (Ketsu): Draw a solution to the dilemma

In particular, the last scene in the climax is very important in manga. The last scene defines the essential problem that appears after the dilemma is resolved. A good last scene leads the reader to understand the essence of the problem, and why the hero wanted to solve the problem. Fig. 3 (shown on the next page) was a manga drawn by students -- the last scene was drawn first.

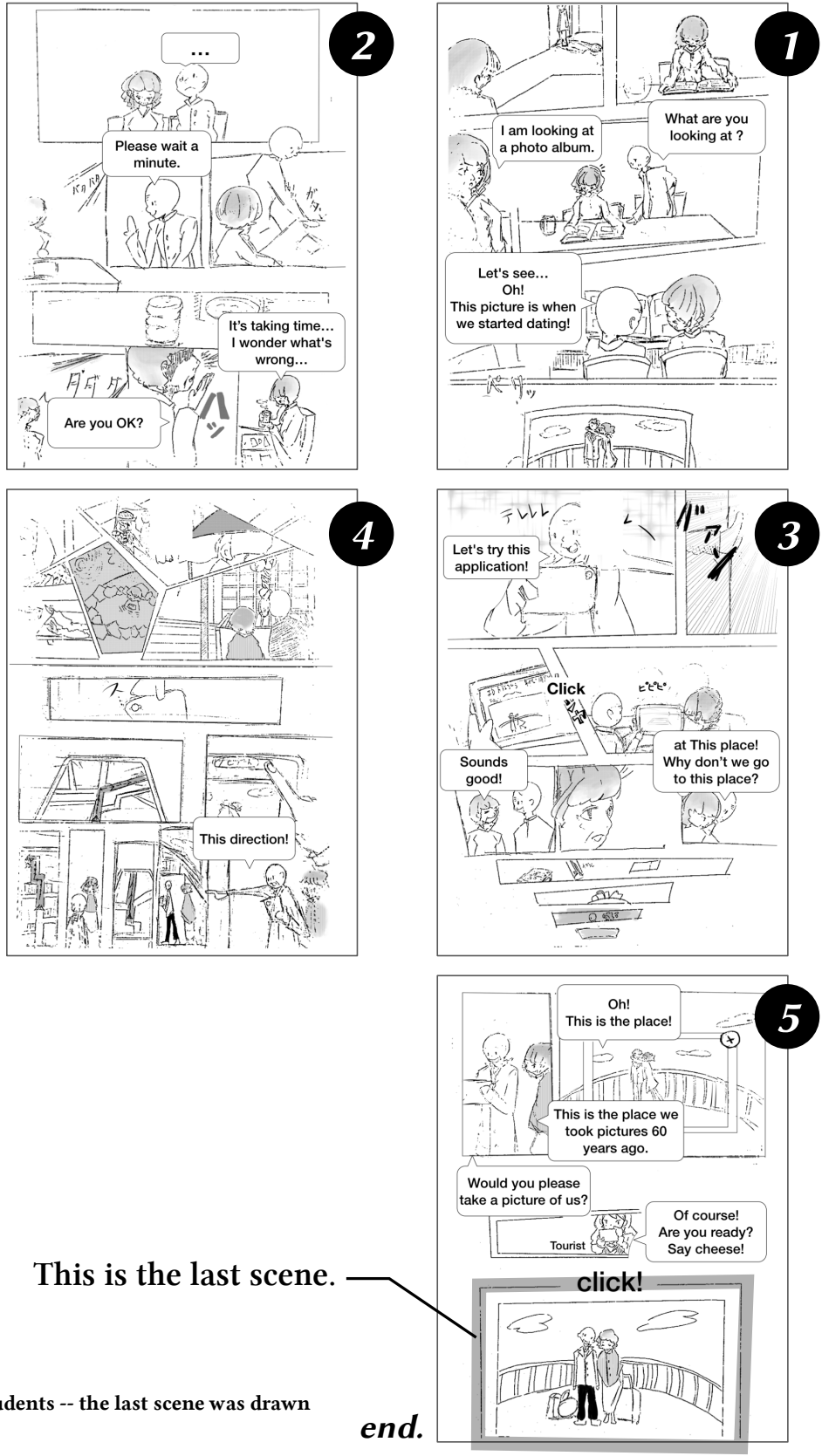

Fig. 3: Manga drawn by students -- the last scene was drawn first.

## 2.3 The Advantages of Manga Techniques

*2.3.1. The Difficulty of Building "The Right Thing."* In his book, "Writing Effective Use Case [10]," Alistair Cockburn presents a framework for Use Case shown in Fig. 4.

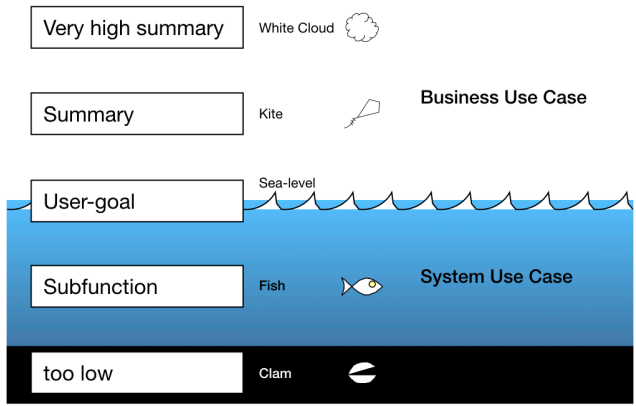

**Fig. 4: Classification of Use Case by Alistair Cockburn.**

According to Cockburn's classification, the "User-goal" is the sea-level. The Business Use Case is defined as above this sea-level, whereas the System Use Case is below the sea-level. Use Case must be structured at each level from the "Very High Summary" (Cloud) to the "Sub-function" (Fish). An engineer should avoid writing a System Use Case in the "Too Low" level, indicated by the Clam icon.

In reality, different levels of Use Cases tend to be mixed. Engineers tend to write scenarios from an implementation perspective that is Clam level, while users tend to explain their requirements in an ad hoc manner. This mixup of levels happens not only in Use Case, but also in other standard methods. As a result, both engineers and users tend to focus on "how" and not "what" to build as a final product.

*2.3.2. Clarifying "Why" with Manga.* As mentioned above, "how" to build a product often overwhelms "what" to build, causing problems not only in the RD processes, but also in the software development processes as a whole. We believe that a reason for this confusion is failing to clearly define "why." The goal for development should be concretely shared between engineers and users as a prerequisite for development. Japanese manga technique provides exactly the right means of communication needed to make a common understanding between users and engineers, specifically by describing "what," "who," and "why." In manga, the user image must be represented with more reality than in Persona. A well represented manga character clearly gives us an understanding of the protagonist's behaviors. The manga provides a "fun" way to come to a common understanding of real problems that users and engineers should solve together.

As shown in Fig. 5, the manga technique removes the Cloud and gives us the "Very high summary", guiding the definition of the Key Summary and Key User-goal. The Key Summary and Key User-goal guide us to understand "What" instead of "How." In other words, Manga gives us a constellation that is a guiding beacon for the voyage. As a result, we can ride the anchor of the ship to ensure good fishing.

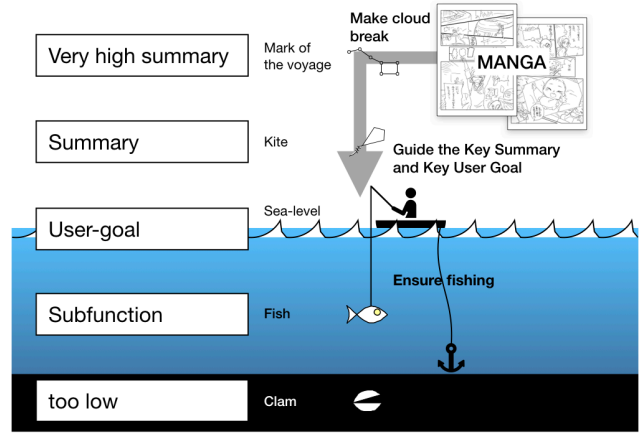

**Fig. 5: Manga gives us a constellation that will be a guiding beacon for the voyage and anchor the ship for good fishing.**

## 3 APPLYING THE MANGA TECHNIQUE TO PBL

### 3.1 Exercise Goals and Design

Our manga PBL exercise aimed at teaching the following topics with hands-on exercises.
- RD process
- Design of services and products by user experience and human-centered design
- Project management
- Process viewpoint by Postmortem
- Facilitation and presentation skills

As mentioned in the introduction, the class was composed of information science students from NAIST and manga majors from OUA. Three to five students from each university mixed in one team to organize three to four teams every year.

As shown in Fig. 6, the manga PBL class uses an RD process organized in five phases. The first phase addresses development of a set of customer requirements. The second phase addresses development of a set of product requirements. The third phase addresses the analysis of user and product requirements needed to define and understand the requirements. The first to third phases are iterated until the verified requirement is defined. Manga techniques are applied in this iterative process. The fourth phase is design prototyping to verify the feasibility of the solution. The last phase is planning a business model. Students make presentations in a real business setting to Panasonic Corporation, a company that supports our exercise.

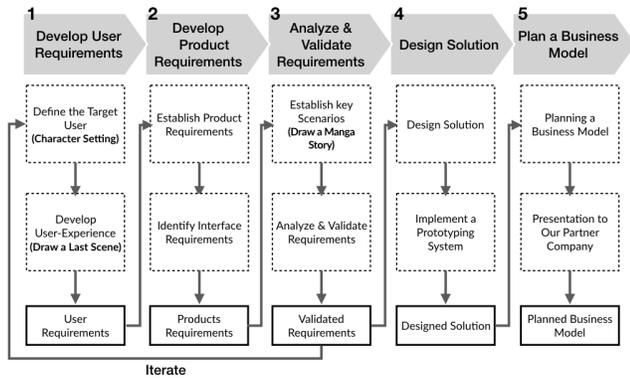

**Fig. 6: The manga PBL process consists of five phases.**

## 3.2 PBL Schedule

The PBL exercise starts in April and ends in December. Fig. 7 shows an overview of the schedule. During the summer, we organize a 3-day workshop. On the last day of this workshop, students make presentations to Panasonic Corporation. At the end of the development, we perform a postmortem. Project postmortem is a process intended to encourage project improvements by determining aspects that were successful and unsuccessful. The class finishes with a final exhibition in December. This exhibition is open to the public, and students present their final products at the exhibition.

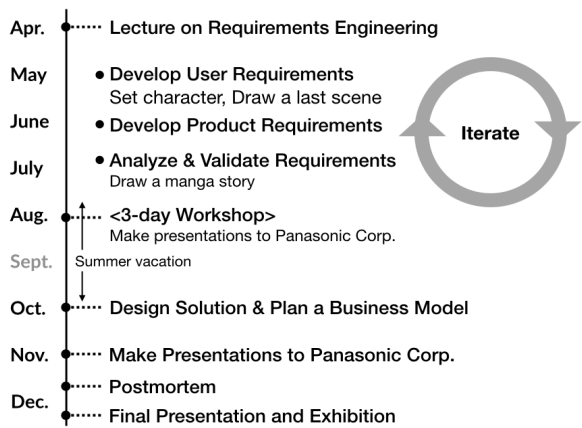

**Fig. 7: PBL starts in April and ends at the December exhibition.**

## 4 EFFECTIVENESS OF APPLYING MANGA TECHNIQUES

As shown in Fig. 8, students drew several manga in their RD process. At this stage of drawing manga, many teams iterate several times. Every year, there is a team which throws out all the ideas and restarts from the very beginning, because we can evaluate the validity of the concept in manga before beginning the development. On the other hand, after validating the concept using manga, the students are focused on design and implementation without hesitation. That is to say that through the process of creating manga, the students were able to clearly define a story from the very high summary goal to the user goal that became the key concept of what to build.

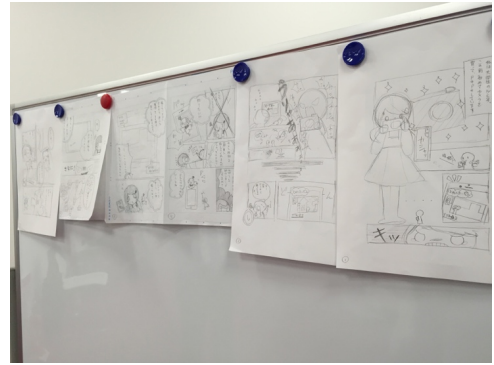

**Fig. 8: Students refine the story by drawing a number of manga.**

Regarding the implementation of interdisciplinary work, we were initially concerned whether art students and information science students could collaborate in one team. However, we saw a very good collaboration between art students and engineering students.

## 5 CONCLUSIONS

By using manga techniques in the RD process, or the phase of defining "what to build" art and information science students were able to communicate well. Every year, all the teams come up with new product ideas, one of which also led to a patent. Recognizing this manga approach as innovative and useful, Panasonic Corporation has pledged to continue their support for the exercise in the new fiscal year.

Our four-year experience with this PBL exercise shows that manga improves communication between users and engineers, and their development goal is clearly shared, so the development phase can proceed smoothly. In Cockburn's terms, the Cloud of the Very High Summary clears up to make the most important User-goal concrete. The RD process is a heuristic activity. In this explorative type process, it is important that users and engineers share the same project goals. We believe the manga technique provides a convenient and reliable tool for sharing the project goals.

In general use of PBL, it may be difficult to have manga majors participate as in this PBL. On the other hand, as shown in Fig. 9, even a rough manga can be sufficiently effective for communication within a team. In fact, there was a team where the information science students drew the manga. We think that the process of expressing requirements with manga is effective even if the manga is not drawn well.

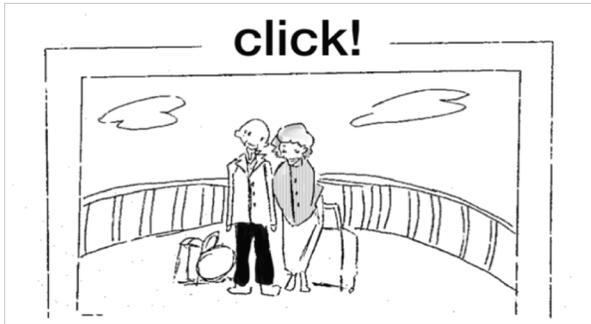

**Fig. 9: Even a rough level manga was sufficiently effective for communication with users and engineers.**